# Yakubovsky scheme to study the 4- and 5-nucleon systems in the case of alpha-state structure with spin-dependent nucleon−nucleon potentials

E. Ahmadi Pouya*, A.A. Rajabi

*Physics Department, Shahrood University of Technology, P. O. Box 3619995161-316, Shahrood, Iran*

**Abstract**

In this project we have investigated the 5-nucleon model system in the picture of the specific alpha-state structure, by extending the Yakubovsky scheme with the inclusion of the spin and isospin degrees of freedom. The Yakubovsky formalism for the 5-nucleon system in the effective alpha-neutron attractive model leads to a set of two coupled equations, based on two relevant alpha-nucleon sub cluster components. To this regard, by switching off the fifth nucleon interactions, the 5-nucleon Yakubovsky equations can be reduced to a typical 4-nucleon problem. To calculate the 5- and 4-nucleon bound state equations, the coupled equations are projected in the momentum space in terms of the Jacobi momenta. In the calculations two different spin-dependent and one spin-independent nucleon−nucleon potential types are dedicated, such as Afnan−Tang $S_3$, Malfliet−Tjon I/III and Volkov potentials, respectively. The some obtained binding energy differences between the 4-nucleon system in the alpha-state and the 5-nucleon system in the case of alpha-nucleon structure demonstrating the effective interaction between the alpha and an attractive neutron. The obtained results for the effective interaction are consistent either for spin-independent and spin-dependent interactions. Also, the binding energy results are in excellent compatibility with the achieved results by other techniques.

*PACS. No:* 21.45.+v; 21.30.Fe; 21.30.-x; 21.10.Dr

*Keywords:* Few-body systems; Forces in hadronic systems and effective interactions; Two-nucleon interactions; Binding energies and masses

## 1. Introduction

Considerable interest has been already shown by both the theorists and the experimentalists in the study of alpha-nucleon interactions, as this process gives insight into nuclear structural problems and also throws some light on the basic two-body interaction. Added interest is also due to the fact that it essentially involves the study of $N = 5$ systems. Incidentally, there is no experimental evidence, so far, for the existence of bound states of five-baryon systems except that of $^5$He. An objection to the use of simple phenomenological potentials for alpha-nucleon ($\alpha N$) scattering arises from the fact that such these potentials allow a bound state, for the five-baryon system. In addition, it is well-known the main importance in the few-body problems are finding an exact solution for the system, also investigating $\alpha N$ interactions, governing on these systems. Therefore, the investigation of scattering and bound states of nuclei interacting via simple and realistic interactions particularly has been always in the center of the interest and description of light $\alpha$-core nuclei, especially considering the effective $\alpha N$ interactions, requires well established approaches to the solution of the non-relativistic Schrodinger equation, as well as investigation of the identity of the effective $\alpha N$ interactions.

* Corresponding author.
*E-mail address:* E.Ahmady.Ph@ut.ac.ir (E. Ahmadi Pouya).

In recent decades, considerable effort has been made to study the effective $\alpha N$ interaction in the scattering analysis, such as multichannel $\alpha N$ and $\alpha\alpha$ interactions [1], bound-state properties of the $^6$He and $^6$Li in a 3-body model, with investigation of $\alpha N$ interactions [2], interactions of $\alpha N$ in an elastic scattering [3], a survey of the $\alpha N$ interaction [4], peripheral $\alpha N$ scattering with $NN$ potential [5] and microscopic calculations of $^5$He with realistic interactions [6] that are very valuable in determining the extent to which nuclei may be successfully described as a system of nucleons. Beside, significant researches has been exerted to obtain accurate ground-state properties of the nuclear bound systems, particularly for $N \geq 4$ with simple and realistic potentials, such as Stochastic Variational Monte Carlo (SVM) technique [7] which, used simplified with two-body interactions, without realistic nuclear forces and the Nonymmetrized Hyperspherical Harmonics (HH) scheme [8] appears to be pretty promising to deal with permutational-symmetry breaking terms in the Hamiltonian. The HH computational schemes are usually based on the partial-wave (PW) method. The SVM technique, however, is made directly using with vectors in the configuration space. The achievements in [7,8] demonstrate that a direct treatment of the 5-nucleon (5N) systems and beyond is now manageable on the todays computers. Therefore, in order to solve the 5N systems, we suggested that a generalizable and old reliable method, the solution within the Yakubovsky equations, is now desirable. Now, after the study of the four- and six-body bound systems in the case $\alpha$-core structure within the powerful Yakubovsky scheme, in a typical partial-wave representation [9] and three-dimensional formalism [10] that the technical expertise has been developed and the very strong increase of computational power just recently achieved allows to study the 5N model system problems in that solution of the Yakubovsky equations for the case of alpha-neutron $(\alpha - n)$ model, to estimate the effective 2-body $\alpha N$ interaction. It is worthwhile to mention that a realistic 5N problem is not allowed for a bound state, however, in order to investigate the effective interactions, namely $\alpha$-particle and an attractive nucleon we make the 5N problem for the case of $\alpha - n$ configuration as a bound system. Also, an objection to the use of simple phenomenological potentials for $\alpha N$ scattering arises from the fact that such these potentials allow a bound state, for the 5N system. Therefore, in order to investigate the effective $\alpha N$ interaction in the specific $\alpha - n$ configuration of the 5N model system, we extend the 5N Yakubovsky equations for $^5$He, extending the applications to spin-dependent particles and in order to calculate the binding energy results, we evaluate the coupled equations in momentum space on the basis of partial-wave decomposition. Next, we have developed a particular representation of the high-sized eigenvalue matrix, which is methodical with respect to the number of components and well suitable for a numerical implementation. In pursuit of this aim, we investigated the accuracy of the calculations regarding the number of grid points and calculate the expectation value of the 5N Hamiltonian operator.

In this article first, we give a brief review of the 5N Yakubovsky formalism by using the standard cluster notation [11] that leads to four coupled equations with four independent components. Next, we select some specific components, where the $\alpha - n$ approximation is valid and work only with two coupled equations in terms of two remaining components. In order to solve the coupled equations we project these components to the corresponding partial-wave basis states based on Jacobi momenta with the inclusion of spin and isospin degrees of freedom. Next, describes details numerical techniques with typical eigenvalue equation form that characterized the dimension of the problem. Next, we introduce the relevant spin-dependent and spin-independent potential models that are used in the numerical calculations and report the binding energy results for 5N $(\alpha - n)$ and 4N ($\alpha$-state) problems, with respect to the regarded obtained from other methods. Finally, we evaluate the expectation value energy for testing the accuracy of the numerical implementations and also the concluding remarks are provided.

## 2. The 5-nucleon Yakubovsky scheme in a sub cluster band

In order to show the specific $\alpha - n$ coupled equations, as well as the anti-symmetrized total state wave function of the effective structure for $^5$He, a brief review of the Yakubovsky equations to the 5N problem using the sub clusters notation [11] is derived. In view of the expectation for the dominant structure of $^5$He, namely considering an inert $\alpha$-state and a bound neutron, the sequential sub clustering can be stopped with 2-body fragments. In a 5-particle system, there are ten different cluster decompositions ($c_4$) having 4 clusters. They are labeled by the only two-body cluster $c_4$ they contain, *e.g.* $c_4 = 12 \equiv 12+3+4+5$. In the following formalism the single particles in sub clusters will no longer be displayed. To the solution of the 5N bound system in Yakubovsky scheme using the sub clusters notation, the idea is to first sum up the pair forces in each sub cluster of 4-body fragments ($c_4$), in a second step among all sub clusters of 3-body fragments ($c_3$), and then in a third step among all sub clusters of 2-body fragments ($c_2$). In the spirit of the usually applied to approximate effective $\alpha - n$ configuration model, it is worked out that formalism ending with 2-body fragments sub clusters. To this end, we start from the following non-relativistic Schrödinger equation for 5N bound system,

$$\left(H_0 + \sum_{c_4}^{10} V_{c_4}\right)\Psi = E\Psi, \tag{1.1}$$

where $H_0$ stands for the kinetic energy operator, $\sum_{c_4}^{10} V_{c_4} \equiv V_{12} + \cdots + V_{45}$ is the summation of the all 4-body fragments sub clusters, that is equivalent with all 2N interactions, and $E$ is the total energy of the 5N system. According to the Yakubovsky scheme, Eq. (1.1) is rewritten into a homogenous (bound state equation) integral equation

$$\Psi = G_0 \sum_{c_4}^{10} V_{c_4}\Psi, \tag{1.2}$$

where $[E - H_0]^{-1}$, stands for the 5N free propagator function and the total wave function $\Psi$, includes the summation of all pair forces as follows

$$\Psi = \sum_{c_4=1}^{10} \psi_{c_4} \equiv \psi_{12} + \psi_{13} + \psi_{14} + \psi_{15} + \psi_{23} + \psi_{24} + \psi_{25} + \psi_{34} + \psi_{35} + \psi_{45}, \tag{1.3}$$

where each pair ($c_4$) having 4-body fragments is sub clusters of some 3-body fragments ($c_3$) as follows

$$\psi_{c_4=12} = \sum_{c_3}^{6} \psi_{c_4,c_3} \equiv \underline{\psi_{12;123}} + \psi_{12;124} + \psi_{12;125} + \underline{\psi_{12;12+34}} + \psi_{12;12+35} + \psi_{12;12+45}, \tag{1.4}$$

where ($c_3$) refers to any 3-body fragments containing the pair restricted to ($c_4 = 12$) and the sum runs over pairs $c_4 \subset c_3$. It means that the sub clusters $c_3$ when broken up lead to the sub clusters $c_4$. Again each 3-body fragments ($c_3$) is sub clusters of some 2-body fragments ($c_2$). The above selected 3-body fragments components are sub clusters of some 2-body fragments,

$$\psi_{c_4=12;c_3=123} = \sum_{c_2}^{3} \psi_{c_4,c_3}^{c_2} \equiv \underline{\psi_{12;123}^{1234}} + \psi_{12;123}^{1235} + \psi_{12;123}^{123+45}, \tag{1.5}$$

$$\psi_{c_4=12;c_3=12+34} = \sum_{c_2}^{3} \psi_{c_4,c_3}^{c_2} \equiv \underline{\psi_{12;12+34}^{1234}} + \psi_{12;12+34}^{125+34} + \psi_{12;12+34}^{12+345}, \tag{1.6}$$

the first two components in the above-mentioned equations, Eqs. (1.5) and (1.6) are related to the very approximate effective $\alpha - n$ specific components. Summing the 4-body fragments in Eq. (1.3), adding 3-body fragments, and corresponding 2-body fragments together, the total 5N wave function $\Psi$, Eq. (1.3) can be obtained with 180 components in which the total wave function can be compacted with suitable permutations. In order to approach the approximate effective $\alpha - n$ structure, the formalism ends up with 2-body fragments, and the relevant Yakubovsky components are selected, where the specific structure approximation is valid. According to the formulation of the 6-body Yakubovsky equations [12], we have implemented the similar derivation to the 5-body system [13], namely after applying the Lippmann–Schwinger equation twice, starting from integral form of the Schrodinger equations, Eq. (1.2), step-by-step and implementing the identity of the nucleons, the 5N bound state formalism finally leads to a set of four coupled equations, Eqs. (1.7)–(1.9) based on four independent components, *i.e.* $\psi_{12;123}^{1234}$, $\psi_{12;12+34}^{1234}$, $\psi_{12;123}^{123+45}$ and $\psi_{12;12+34}^{125+34} + \psi_{12;12+34}^{12+345}$. They are all independent 2-body fragments sub clusters of the Yakubovsky special wave functions. Also more discussions in details could be found in Ref. [13]. Therefore, the 5N Yakubovsky equations in terms of the above-mentioned four independent components yields

$$\begin{pmatrix} \psi_{12;123}^{1234} \\ \psi_{12;12+34}^{1234} \end{pmatrix} = G_0 \begin{pmatrix} \mathcal{T}^{123}(-P_{34}) & \mathcal{T}^{123} \\ \mathcal{T}^{12+34}(1-P_{34}) & 0 \end{pmatrix} \times \left[\begin{pmatrix} -P_{45}\psi_{12,123}^{1234} + \psi_{12,123}^{123+45} \\ \psi_{12;12+34}^{125+34} + \psi_{12;12+34}^{12+345} \end{pmatrix} + \begin{pmatrix} \psi_{12;123}^{1234} \\ \psi_{12;12+34}^{1234} \end{pmatrix}\right], \tag{1.7}$$

$$\psi^{123+45}_{12,123} = G_0\mathscr{T}^{123}(-P_{35})\Big(\Big(\psi^{125+34}_{12,12+34} + \psi^{12+345}_{12,12+34}\Big) + \psi^{1234}_{12;12+34}\Big), \tag{1.8}$$

$$\psi^{125+34}_{12,12+34} + \psi^{12+345}_{12,12+34} = G_0\mathscr{T}^{12+34}\Big((-P_{35} - P_{45})\Big(\psi^{125+34}_{12,12+34} + \psi^{12+345}_{12,12+34}\Big) - P_{45}(1-P_{34})\psi^{1234}_{12;12+34} - P_{35}\Big((1-P_{34})\psi^{1234}_{12;123} + \psi^{123+45}_{12,123}\Big)\Big), \tag{1.9}$$

regarding the sub cluster underlying the four components only two of them, that is $\psi^{1234}_{12;123}$ and $\psi^{1234}_{12;12+34}$, are related to the very approximative 2-body fragments configuration model of an inert $\alpha$-state and an attractive neutron, where the approximation of $\alpha$-particle and attractive neutron is valid. The component $\psi^{123+45}_{12,123}$ mentions to an inert 3N together with a 2N sub cluster for the 5N system. The linear combinations $(\psi^{125+34}_{12,12+34} + \psi^{12+345}_{12,12+34})$ refers again to an inert 3N together with a 2N sub system, where the underlying fragmentation related to 2-body fragments differ from $\psi^{123+45}_{12,123}$. Hence, it is obvious that the only first two components, $\psi^{1234}_{12;123}$ and $\psi^{1234}_{12;12+34}$, are related to the very approximate effective 2-body fragments model of an inert $\alpha - n$ and other components will not be taken into account. Therefore, to establish the expression for the total state of the $\alpha - n$ structure, now it is desirable to switch off irrelevant components from the total state wave function $\Psi$. Therefore, the $\alpha - n$ structure wave function with 90 components yields

$$\Psi_{\alpha-n} = [1 - P_{23} - P_{24} - P_{13} - P_{14} + P_{13}P_{24}]\Big[(1-P_{34})\psi^{1234}_{12;123} + \psi^{1234}_{12;12+34}\Big] - [1 - P_{23} - P_{24} - P_{13} - P_{14} + P_{13}P_{24}] \times \Big[((1-P_{34})(P_{45}+P_{35}))\psi^{1234}_{12;123} + (P_{45}+P_{35})\psi^{1234}_{12;12+34}\Big] - [P_{25} + P_{15} - P_{13}P_{25} - P_{14}P_{25}] \times \Big[(1-P_{34})\psi^{1234}_{12;123} + \psi^{1234}_{12;12+34} - ((1-P_{34})(P_{45}+P_{35}))\psi^{1234}_{12;123} - (P_{45}+P_{35})\psi^{1234}_{12;12+34}\Big]. \tag{1.10}$$

In order to consider the 3N forces, in addition to the 2N interactions in the Hamiltonian form in Eq. (1.1) and to calculate the expectation value of the Hamiltonian operator, the above-mentioned total wave functions have to be used. Clearly, the above-mentioned total wave function is anti-symmetrized following the Pauli principle, *i.e.* $\Psi_{\alpha-n} = -P_{ij}\Psi_{\alpha-n}$. It is well-known the first piece with 18 components has exactly the form of the total wave function of a 4N bound state problem [14].

$$\Psi_\alpha = [1 - P_{23} - P_{24} - P_{13} - P_{14} + P_{13}P_{24}]\Big[(1-P_{34})\psi^{1234}_{12;123} + \psi^{1234}_{12;12+34}\Big]. \tag{1.11}$$

Hence, for 5N bound system, the two above-mentioned components, as a physical 4N problem in Eq. (1.11), namely $\psi^{1234}_{12;123}$ and $\psi^{1234}_{12;12+34}$, not only do refer to the contribution of the nucleons 1—4, but also have contribution of nucleons 5. In the next step the specific 5N coupled equations as an $\alpha - n$ system are represented in momentum space in terms of the Jacobi momenta.

## 3. Momentum space representation

In this section, first we explain that why we choice the specific components, namely related component with $\alpha - n$ configurations. Though, it is worthwhile to mention that for full solution of the general case of 5N system, we need modern super-computers with grid parallel treatment. Therefore, we should consider all of the four coupled equations, because all four independent Yakubovsky components are equally important in general case [13]. But here, we would like to study the bound state of 5N system for the case of specific $\alpha - n$ structure. Therefore, we choose the first two components, and the other components will not be taken into account in the specific $\alpha - n$ structure. Also, according to Fig. 1, the effective interaction of $\alpha$-particle is governor in the remained components (see Fig. 1 and compare with Figs. 1 and 2 in Ref. [14]). For approximating effective interaction as a 5N effective $\alpha - n$ structure, we selected $\psi^{1234}_{12;123}$ and $\psi^{1234}_{12;12+34}$. By

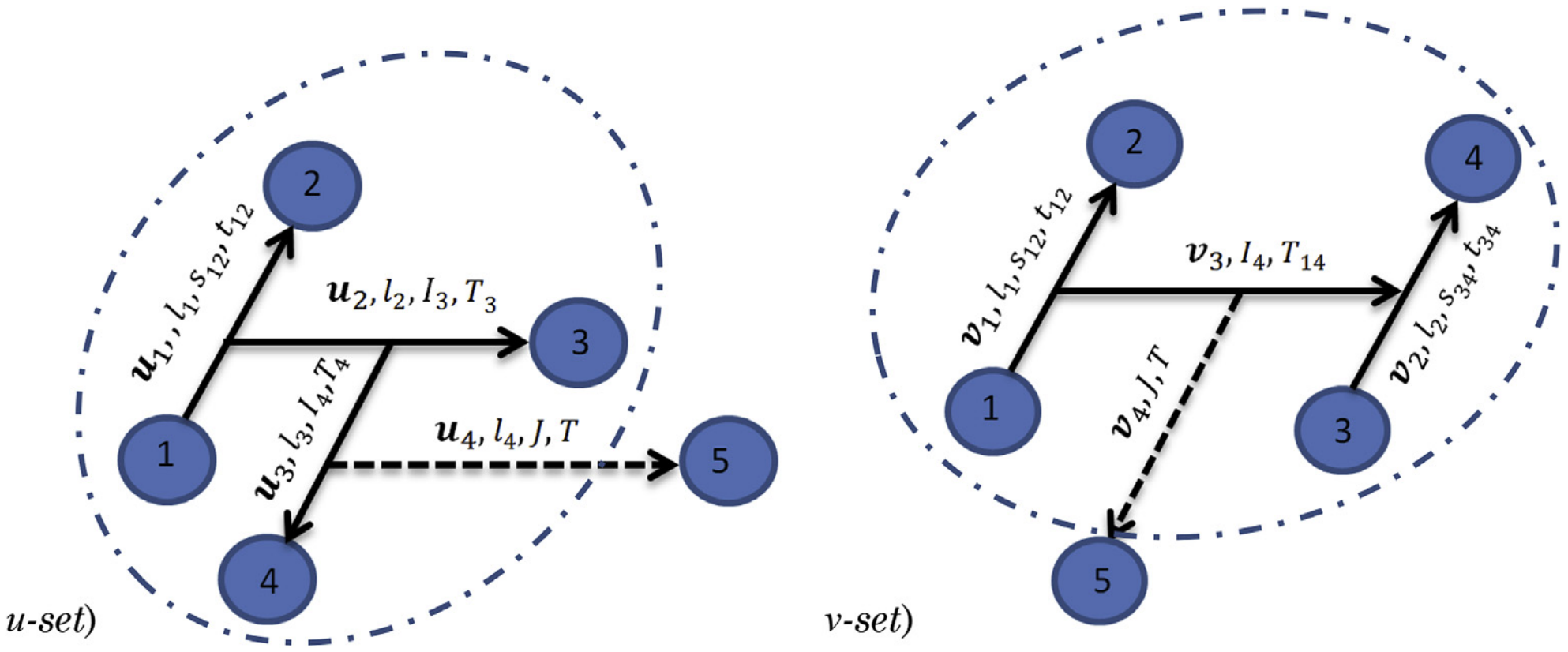

Fig. 1. Schematic configurations of the 5N system as an $\alpha - n$ structure in Jacobi coordinates. The $u$- and $v$-sets are related to $\psi^{1234}_{12;123}$ and $\psi^{1234}_{12;12+34}$, respectively. In the bound circle the $\alpha$-state plays as a 4N subsystem.

these considerations, the remaining two first coupled equations, Eq. (1.7), lead to

$$\begin{pmatrix}\psi^{1234}_{12;123}\\ \psi^{1234}_{12;12+34}\end{pmatrix} = G_0\begin{pmatrix}\mathscr{T}^{123}(-P_{34}) & \mathscr{T}^{123}\\ \mathscr{T}^{12+34}(1-P_{34}) & 0\end{pmatrix} \times\left[\begin{pmatrix}(1-P_{45})\psi^{1234}_{12;123}\\ \psi^{1234}_{12;12+34}\end{pmatrix}\right]. \tag{2.1}$$

It is well-known that nuclear systems should be treated in the fermionic approaches, and the Pauli principle is taken into account, even for spinless particles. Also here, we describe the spin and isospin degrees of freedom in corresponding basis states and study the 5N system in the $L = 1$ states, Therefore, symmetrizing the special wave function that justifies this 5N system is equivalent to a one neutron around an inert $\alpha$-state with isospin $T = 1/2$ and $m_T = -1/2$ (one defines the magnetic isospin quantum numbers of the neutron $-1/2$), wherever the total angular momentum quantum numbers of $^5$He ground state is $(L = 1, S = 1/2, T = 1/2)j^\pi = 3^-/2$ (see Tables 2 and 3 in Section 4). In order to describe the corresponding momentum basis states, for two specific components of the effective $\alpha - n$ configurations, the standard Jacobi momenta with the angular momentum, spin and isospin quantum numbers are described in Fig. 1.

According to the above-mentioned specific configurations, Fig. 1, after removing the contribution of the fifth nucleon in the 5N coupled equations, Eq. (2.1), namely $P_{45}\equiv 0$, the 5N system leads to a typical 4N problem [14] as follows

$$\psi^{1234}_{12;123} = -G_0\mathscr{T}^{123}P_{34}\psi^{1234}_{12;123} + G_0\mathscr{T}^{123}\psi^{1234}_{12;12+34}, \tag{2.2}$$

$$\psi^{1234}_{12;12+34} = G_0\mathscr{T}^{12+34}(1-P_{34})\psi^{1234}_{12;123}, \tag{2.3}$$

such reduction confirms that the Yakubovsky formalism for 5N bound system in the specific structure is a valid approximation and a powerful approach to describe the effective $\alpha - n$ structure. Therefore, in addition to the bound state calculations of the 5N system, we also calculate the 4N bound system for comparison. In order to describe the effective Yakubovsky components of 5N system in the case of $\alpha - n$ structure, in momentum space representation, we introduce standard Jacobi momenta corresponding to the first two components. For $\psi^{1234}_{12;123}$ in terms of the first configuration in Fig. 1, sets corresponding to $u$-set chains

$$\begin{pmatrix}\boldsymbol{u}_1\\ \boldsymbol{u}_2\\ \boldsymbol{u}_3\\ \boldsymbol{u}_4\\ K\end{pmatrix} = \begin{pmatrix}1/2 & -1/2 & 0 & 0 & 0\\ -1/3 & -1/3 & 2/3 & 0 & 0\\ -1/4 & -1/4 & -1/4 & 3/4 & 0\\ -1/5 & -1/5 & -1/5 & -1/5 & 4/5\\ 1 & 1 & 1 & 1 & 1\end{pmatrix}\begin{pmatrix}\boldsymbol{k}_1\\ \boldsymbol{k}_2\\ \boldsymbol{k}_3\\ \boldsymbol{k}_4\\ \boldsymbol{k}_5\end{pmatrix}. \tag{2.4}$$

In the non-relativistic case we may express the kinetic energy operator by two equivalent forms. The inverse form of the above-mentioned transfer matrix is used for representation of the kinetic energy in terms of $u$-set Jacobi momenta.

$$H^u_0 = \sum_{i=1}^{5}\frac{k_i^2}{2m} = \frac{u_1^2}{m} + \frac{3}{4}\frac{u_2^2}{m} + \frac{2}{3}\frac{u_3^2}{m} + \frac{5}{8}\frac{u_4^2}{m}, \tag{2.5}$$

Table 1
The number of partial-wave channels in low-laying states contributing to both kinds of the 5N Jacobi coordinates for $J = 3^-/2$. $N^S_{\gamma_u}$ ($N^T_{\gamma_u}$) and $N^S_{\gamma_v}$ ($N^T_{\gamma_v}$) are the number of the spin (isospin) states respectively, for both-set Jacobi coordinates in Fig. 1. For $u$- and $v$-set parts the total spin and isospin are restricted to $J = 3/2$ and $T = 1/2$, respectively. In this regard, the results of part (I) and (II) are given according to the spin and isospin of the $^5$He state, where $s_5$ and $t_5$ are fixed to be one-half.

*(I): u-set Jacobi partition*

| $\left(s_{12}\frac{1}{2}\right)I_3\left(I_3\frac{1}{2}\right)I_4(I_4s_5)J$ | $S=\frac{1}{2}$ | $\left(t_{12}\frac{1}{2}\right)t_3\left(t_3\frac{1}{2}\right)T_4(T_4t_5)T$ | $T=\frac{1}{2}$ |
|---|---|---|---|
| $\left(0\frac{1}{2}\right)\frac{1}{2}\left(\frac{1}{2}\frac{1}{2}\right)0\left(0\frac{1}{2}\right)\frac{1}{2}$ | 1 | $\left(0\frac{1}{2}\right)\frac{1}{2}\left(\frac{1}{2}\frac{1}{2}\right)0\left(0\frac{1}{2}\right)\frac{1}{2}$ | 1 |
| $\left(0\frac{1}{2}\right)\frac{1}{2}\left(\frac{1}{2}\frac{1}{2}\right)1\left(1\frac{1}{2}\right)\frac{3}{2}$ | 0 | $\left(0\frac{1}{2}\right)\frac{1}{2}\left(\frac{1}{2}\frac{1}{2}\right)1\left(1\frac{1}{2}\right)\frac{3}{2}$ | 0 |
| $\left(1\frac{1}{2}\right)\frac{1}{2}\left(\frac{1}{2}\frac{1}{2}\right)0\left(0\frac{1}{2}\right)\frac{1}{2}$ | 1 | $\left(1\frac{1}{2}\right)\frac{1}{2}\left(\frac{1}{2}\frac{1}{2}\right)0\left(0\frac{1}{2}\right)\frac{1}{2}$ | 1 |
| $\left(1\frac{1}{2}\right)\frac{1}{2}\left(\frac{1}{2}\frac{1}{2}\right)1\left(1\frac{1}{2}\right)\frac{3}{2}$ | 0 | $\left(1\frac{1}{2}\right)\frac{1}{2}\left(\frac{1}{2}\frac{1}{2}\right)1\left(1\frac{1}{2}\right)\frac{3}{2}$ | 0 |
| $\left(1\frac{1}{2}\right)\frac{3}{2}\left(\frac{3}{2}\frac{1}{2}\right)1\left(1\frac{1}{2}\right)\frac{3}{2}$ | 0 | $\left(1\frac{1}{2}\right)\frac{3}{2}\left(\frac{3}{2}\frac{1}{2}\right)1\left(1\frac{1}{2}\right)\frac{3}{2}$ | 0 |
| $\left(1\frac{1}{2}\right)\frac{3}{2}\left(\frac{3}{2}\frac{1}{2}\right)2\left(2\frac{1}{2}\right)\frac{5}{2}$ | 0 | $\left(1\frac{1}{2}\right)\frac{3}{2}\left(\frac{3}{2}\frac{1}{2}\right)2\left(2\frac{1}{2}\right)\frac{5}{2}$ | 0 |
| $N^S_{\gamma_u}$ | 2 | $N^T_{\gamma_u}$ | 2 |

*(II): v-set Jacobi partitions*

| $(s_{12}s_{34})S\left(S\frac{1}{2}\right)J$ | $S=\frac{1}{2}$ | $(t_{12}t_{34})t_4\left(t_4\frac{1}{2}\right)T$ | $T=\frac{1}{2}$ |
|---|---|---|---|
| $(00)0\left(0\frac{1}{2}\right)0$ | 1 | $(00)0\left(0\frac{1}{2}\right)\frac{1}{2}$ | 1 |
| $(01)1\left(1\frac{1}{2}\right)\frac{3}{2}$ | 0 | $(01)1\left(1\frac{1}{2}\right)\frac{3}{2}$ | 0 |
| $(10)1\left(1\frac{1}{2}\right)\frac{3}{2}$ | 0 | $(10)1\left(1\frac{1}{2}\right)\frac{3}{2}$ | 0 |
| $(11)0\left(0\frac{1}{2}\right)\frac{1}{2}$ | 1 | $(11)0\left(0\frac{1}{2}\right)\frac{1}{2}$ | 1 |
| $(11)2\left(2\frac{1}{2}\right)\frac{5}{2}$ | 0 | $(11)2\left(2\frac{1}{2}\right)\frac{5}{2}$ | 0 |
| $N^S_{\gamma_v}$ | 2 | $N^T_{\gamma_v}$ | 2 |

Table 2
The number of coupled equations for the 5N in a partial-wave representation according to spin-isospin states $(S-T)$ that we have taken into account. $N_{S/T}$ is the total number of equations, where the $N_{\gamma_u}$ and $N_{\gamma_v}$ are the number of $u$-set and $v$-set channels correspondingly.

| $(S-T)$ | $N_{\gamma_u} = N^S_{\gamma_u} \times N^T_{\gamma_u}$ | $N_{\gamma_v} = N^S_{\gamma_v} \times N^T_{\gamma_v}$ | $N_{S/T} = N_{\gamma_u} + N_{\gamma_v}$ |
|---|---|---|---|
| $\left(\frac{1}{2}-\frac{1}{2}\right)$ | 4 | 4 | 8 |

Table 3
The four- and five-nucleon binding energies for spin-independent Volkov potential.

| Method | $E_{4N}$ (MeV) | $E_{5N}$ (MeV) |
|---|---|---|
| VAR [21] | −30.317 | |
| HH [22] | −30.406 | −42.383 |
| SVM [7] | −30.42 | −43.00 |
| HH [23,24] | −30.420 | −43.032 |
| This work | −30.39 | −44.02 |

where $\boldsymbol{k}_i$ is individual particle momentum in the center of mass situation ($\boldsymbol{K} \equiv \sum \boldsymbol{k}_i = 0$), that described by relative Jacobi momenta ${}^i\boldsymbol{u}_i$ $(i = 1, 2, 3, 4)$. Similarly, to $\psi^{1234}_{12;12+34}$ in terms of the second configuration in Fig. 1, belongs

$$\begin{pmatrix} \boldsymbol{v}_1 \\ \boldsymbol{v}_2 \\ \boldsymbol{v}_3 \\ \boldsymbol{v}_4 \\ \boldsymbol{K} \end{pmatrix} = \begin{pmatrix} 1/2 & -1/2 & 0 & 0 & 0 \\ 0 & 0 & 1/2 & -1/2 & 0 \\ 1/2 & 1/2 & -1/2 & -1/2 & 0 \\ -1/5 & -1/5 & -1/5 & -1/5 & 4/5 \\ 1 & 1 & 1 & 1 & 1 \end{pmatrix} \begin{pmatrix} \boldsymbol{k}_1 \\ \boldsymbol{k}_2 \\ \boldsymbol{k}_3 \\ \boldsymbol{k}_4 \\ \boldsymbol{k}_5 \end{pmatrix}. \tag{2.6}$$

Correspondingly, the kinetic energy in terms of $v$-set Jacobi momenta, are given as

$$H_0^v = \sum_{i=1}^{5} \frac{k_i^2}{2m} = \frac{v_1^2}{m} + \frac{v_2^2}{m} + \frac{1}{2}\frac{v_3^2}{m} + \frac{5}{8}\frac{v_4^2}{m}. \tag{2.7}$$

Now, we can introduce the basis states corresponding to the two specific independent components. According to

Fig. 1 the partial-wave representation of the basis states suitable for $\psi^{1234}_{12;123}$, are given as

$$|u\rangle \equiv |u_1u_2u_3u_4;\gamma_u\rangle \equiv \Big|u_1u_2u_3u_4;(l_1s_{12})j_1\left(l_2\frac{1}{2}\right)j_2(j_1j_2)I_3\left(l_3\frac{1}{2}\right)j_3\left(l_4\frac{1}{2}\right)j_4(j_3j_4)I_4(I_3I_4)J,M_J\rangle \otimes \Big|\left(t_{12}\frac{1}{2}\right)T_3\left(T_3\frac{1}{2}\right)T_4\left(T_4\frac{1}{2}\right)T,M_T\rangle, \quad (2.8)$$

here the orbital angular momenta $l_i$ go with the $\boldsymbol{u}_i$, $s_{ij}$ are 2N spins for nucleons $ij$, $j_i$ are total 1 and 2N angular momenta coupled out of orbital and spin angular momenta, $I_3$ and $I_4$ are total 3N and 4N angular momenta, $j_4$ the total angular momentum of 5N and finally $I_3$ and $I_4$ are coupled to $J$, the conserved total 5N angular momentum. The second state in Eq. (2.8) refers to isospin in an obvious manner. Correspondingly, the basis states for $\psi^{1234}_{12;12+34}$ are given as

$$|v\rangle \equiv |v_1v_2v_3v_4;\gamma_v\rangle \equiv \Big|v_1v_2v_3v_4;(l_1s_{12})j_1(l_2s_{34})j_2(j_1j_2)S(l_3S)I_4\left(l_4\frac{1}{2}\right)j_4(I_4j_4)J,M_J\rangle \otimes \Big|(t_{12}t_{34})T_4\left(T_4\frac{1}{2}\right)T,M_T\rangle, \quad (2.9)$$

where the orbital angular momenta $l_i$ go with the $\boldsymbol{v}_i$, $s_{ij}$ are 2N spins for nucleons $ij$, $S$ are total 4N angular momenta, $j_4$ the total angular momentum of 5N and finally $I_4$ and $j_5$ are coupled to $J$, the conserved total 5N angular momentum. The isospin coupling should be obvious. For the sake of simplicity we switch off orbital angular momentum quantum numbers, though, in the numerical techniques we describe dependent on angular grid points by choosing relevant coordinate systems. By choosing the coordinate systems the orbital angular momentums of the 5N system are automatically taken into account. Clearly, both the above-mentioned basis states are orthonormal in Hilbert space.

$$\int V^2 DV\big|V_1V_2V_3V_4;\gamma_{S,T}\rangle\langle V_1V_2V_3V_4;\gamma_{S,T}\big| = 1;\ V^2DV \equiv V_1^2dV_1V_2^2dV_2V_3^2dV_3V_4^2dV_4, \quad (2.10)$$

where $V_i$ indicates each one of $u_i$ and $v_i$ magnitude of vectors. In addition to the consideration of the spin and isospin effects to the above-mentioned basis states, we can make usage of that basis states without spin and isospin effects, *i.e.* $\gamma_u = \gamma_v = 0$, and here correspondingly we apply the spin-independent interactions. But, because of dealing with fermions, the Pauli principle is taken into account, and by selecting the suitable coordinate systems, we grid to consider the 5N system in $L = 1$. This scenario can be acceptable in few-body problems in nuclear physics. Obviously, the both basis are complete in the 5N Hilbert space. Now, we can project the obtained coupled equations, Eq. (2.1) in corresponding partial-wave basis states. Let us now represent the coupled equations, Eq. (2.1) by inserting the convenient relations, Eq. (2.10), between the permutation operators, it results

$$\begin{aligned}\langle u;\gamma_u\big|\psi^{1234}_{12;123}\rangle &= \int u'^2du'\int u''^2du''\langle u;\gamma_u|G_0\mathscr{T}^{123}|u';\gamma_{u'}\rangle\langle u';\gamma_{u'}|P_{34}P_{45}|u'';\gamma_{u''}\rangle\langle u'';\gamma_{u''}\big|\psi^{1234}_{12;123}\rangle\\ &-\int u'^2du'\int u''^2du''\langle u;\gamma_u|G_0\mathscr{T}^{123}|u';\gamma_{u'}\rangle\langle u';\gamma_{u'}|P_{34}|u'';\gamma_{u''}\rangle\langle u'';\gamma_{u''}\big|\psi^{1234}_{12;123}\rangle\\ &+\int u'^2du'\int v'^2dv'\langle u;\gamma_u|G_0\mathscr{T}^{123}|u';\gamma_{u'}\rangle\langle u';\gamma_{u'}|v';\gamma_{v'}\rangle\langle v';\gamma_{v'}\big|\psi^{1234}_{12;12+34}\rangle,\end{aligned} \quad (2.11)$$

$$\begin{aligned}\langle v;\gamma_v|\psi^{1234}_{12;12+34}\rangle = & \int v'^2 dv' \int u'^2 du' \langle v;\gamma_v|G_0\mathcal{T}^{12+34}|v';\gamma_{v'}\rangle\langle v';\gamma_{v'}|u';\gamma_{u'}\rangle\langle u';\gamma_{u'}|\psi^{1234}_{12;123}\rangle \\ & - \int v'^2 dv' \int u'^2 du' \langle v;\gamma_v|G_0\mathcal{T}^{12+34}|v';\gamma_{v'}\rangle\langle v';\gamma_{v'}|P_{45}|u';\gamma_{u'}\rangle\langle u';\gamma_{u'}|\psi^{1234}_{12;123}\rangle \\ & - \int v'^2 dv' \int u'^2 du' \langle v;\gamma_v|G_0\mathcal{T}^{12+34}|v';\gamma_{v'}\rangle\langle v';\gamma_{v'}|P_{34}|u';\gamma_{u'}\rangle\langle u';\gamma_{u'}|\psi^{1234}_{12;123}\rangle \\ & + \int v'^2 dv' \int u'^2 du' \langle v;\gamma_v|G_0\mathcal{T}^{12+34}|v';\gamma_{v'}\rangle\langle v';\gamma_{v'}|P_{34}P_{45}|u';\gamma_{u'}\rangle\langle u';\gamma_{u'}|\psi^{1234}_{12;123}\rangle, \end{aligned} \quad (2.12)$$

the various terms appearing in the right hand side of Eqs. (2.11) and (2.12) are explicitly evaluated in Appendix. After evaluation of each term in the right hand side of Eqs. (2.11) and (2.12) and using the Delta-function completeness and diagonal properties of the Hilbert space in a partial-wave representation, as well as evaluation of adequate Clash—Gordon coefficients the coupled integral equations are the starting point for the numerical calculations as an eigenvalue equation form, using an iteration method [15].

## 4. Numerical techniques

After evaluation of each term in the above-mentioned coupled integral equations in the standard partial-wave representation, the obtained equations are the starting point for numerical calculations as an eigenvalue equation form. In order to reduce the high sized of the problem, first we choose a suitable coordinate system. We draw with the third vector $\boldsymbol{V}_3$ with parallel to $z$-axis, the second vector $\boldsymbol{V}_2$ in the $x$—$z$ plane and the first vector $\boldsymbol{V}_1$ and fourth vector $\boldsymbol{V}_4$ are arbitrary in the space. Therefore, we need ten variables to uniquely specify the geometry of the four vectors $\boldsymbol{V}_i$ $(i = 1, \ldots, 4)$ with three spherical and two azimuthal angles variables, and last spin variable. According to the selected coordinate system, Fig. 2, the independent angle variables between the Jacobi momenta in both-sets are therefore given by

$$u_{31}\equiv\cos(\boldsymbol{u}_3,\boldsymbol{u}_1)=\cos\vartheta_1 \qquad v_{31}\equiv\cos(\boldsymbol{v}_3,\boldsymbol{v}_1)=\cos\theta_1$$
$$u_{32}\equiv\cos(\boldsymbol{u}_3,\boldsymbol{u}_2)=\cos\vartheta_2 \qquad v_{32}\equiv\cos(\boldsymbol{v}_3,\boldsymbol{v}_2)=\cos\theta_2$$
$$u_{34}\equiv\cos(\boldsymbol{u}_3,\boldsymbol{u}_4)=\cos\vartheta_4 \qquad v_{34}\equiv\cos(\boldsymbol{v}_3,\boldsymbol{v}_4)=\cos\theta_4$$
$$u_{21}\equiv\cos(\boldsymbol{u}_2,\boldsymbol{u}_1)=\widehat{\boldsymbol{u}}_2\cdot\widehat{\boldsymbol{u}}_1=u_2u_1+\sqrt{1-u_2^2}\sqrt{1-u_1^2}\cos(\varphi_1) \qquad v_{21}\equiv\cos(\boldsymbol{v}_2,\boldsymbol{v}_1)=\widehat{\boldsymbol{v}}_2\cdot\widehat{\boldsymbol{v}}_1=v_2v_1+\sqrt{1-v_2^2}\sqrt{1-v_1^2}\cos(\phi_1)$$
$$u_{24}\equiv\cos(\boldsymbol{u}_2,\boldsymbol{u}_4)=\widehat{\boldsymbol{u}}_2\cdot\widehat{\boldsymbol{u}}_4=u_2u_4+\sqrt{1-u_2^2}\sqrt{1-u_4^2}\cos(\varphi_4) \qquad v_{24}\equiv\cos(\boldsymbol{v}_2,\boldsymbol{v}_4)=\widehat{\boldsymbol{v}}_2\cdot\widehat{\boldsymbol{v}}_4=v_2v_4+\sqrt{1-v_2^2}\sqrt{1-v_4^2}\cos(\phi_4) \tag{3.1}$$

in the above equations the $\widehat{u}_i$ ($\widehat{v}_i$) is individual unit vector of $\boldsymbol{u}_i$ ($\boldsymbol{v}_i$). By this selection the numerical challenges of the two- and three-fold integral terms, could be resolved, because we implement the same integral-factor based on new angle factors. Therefore, we replace the old integral factors, Eqs. (A.17) and (A.26), with new factors, as follows:

$$\int_{-1}^{1} du_{32}\int_{-1}^{1} du_{42}\int_{-1}^{1} du_{34}\equiv\int_{-1}^{1} d\cos\theta_2\int_{0}^{2\pi} d\varphi_2\int_{-1}^{1} d\cos\theta_4, \tag{3.2}$$

$$V_{\text{triplet}}(r) = 1000.0\exp\left[-3.0r^2\right] - 326.7\exp\left[-1.05r^2\right] - 43.0\exp\left[-0.6r^2\right]\ [\text{MeV}],$$
$$V_{\text{Singlet}}(r) = 1000.0\exp\left[-3.0r^2\right] - 166.0\exp\left[-0.8r^2\right] - 23.0\exp\left[-0.4r^2\right]\ [\text{MeV}], \tag{3.5}$$

$$\int_{-1}^{1} dv_{32'}\int_{-1}^{1} dv_{42'}\int_{-1}^{1} dv_{34}\equiv\int_{-1}^{1} d\cos\vartheta_2\int_{0}^{2\pi} d\varphi_2\int_{-1}^{1} d\cos\vartheta_4. \tag{3.3}$$

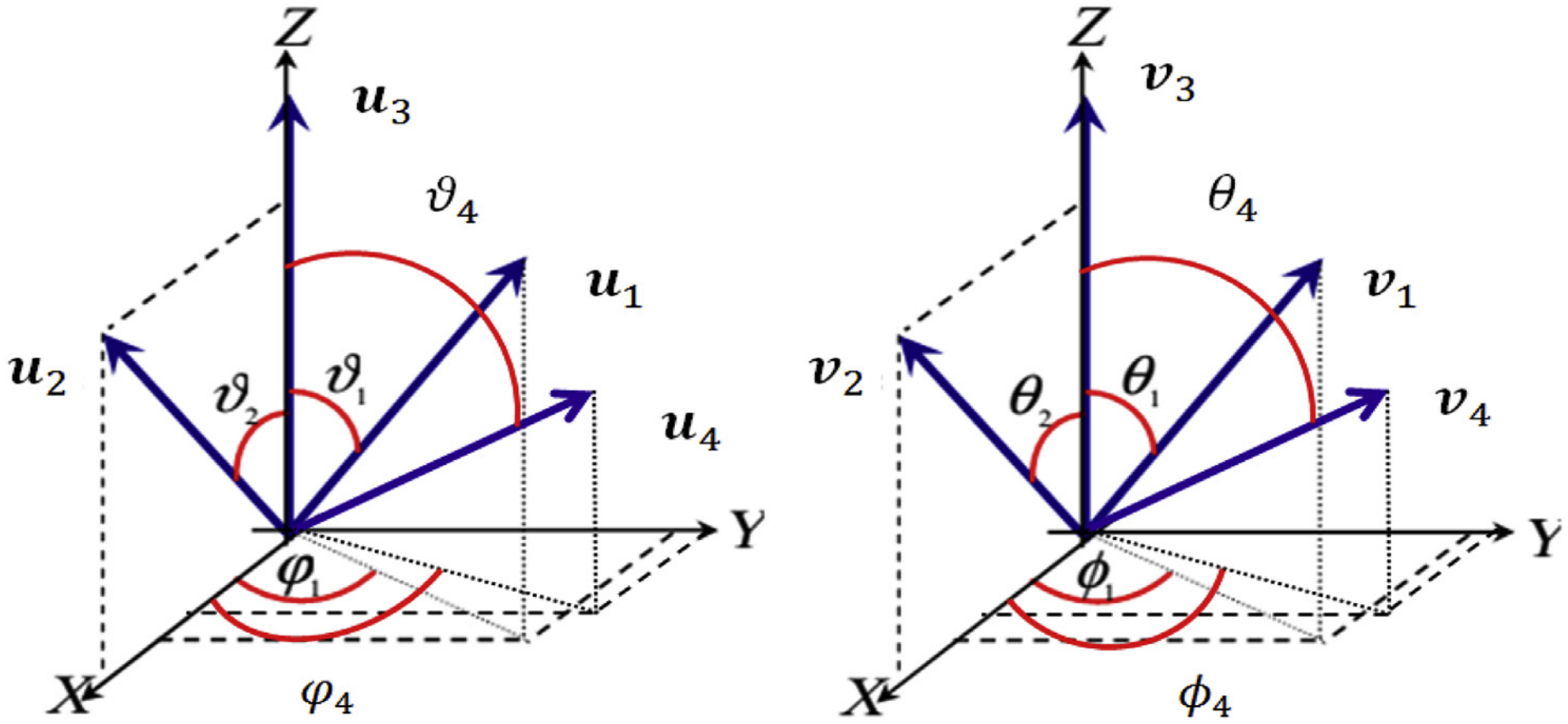

Fig. 2. These coordinates represent the geometry of both sets Jacobi momentum vectors $\boldsymbol{u}_i$ and $\boldsymbol{v}_i$ corresponding to the Yakubovsky components.

We replace the continuous variables in the numerical treatment by a dependence on certain distinct values by using the Gauss–Legendre discretization. The eigenvalue equation is solved by the iteration method. We use a Lanczos-like scheme that is efficient for nuclear few-body problems [9,10,14]. The energy is varied such that one reaches eigenvalue to be one. Since the evaluated coupled integral equations require a very large number of interpolations, we use the cubic Hermit splines of Ref. [15] for its accuracy and high computational speed. More discussions in details for numerical techniques can be found in Refs. [9,16].

In order to be able to match our calculations with obtained by other techniques, we have used the relevant spin-independent and also spin-dependent potential models ($r$ is always in $fm\equiv10^{-15}$ m). The nucleon–nucleon potential model applied in the calculations are defined as

I) Spin-independent Gauss-type Volkov potential [17].

$$V(r) = 144.86\exp\left[-1.487r^2\right] - 83.34\exp\left[-0.3906r^2\right]\ [\text{MeV}], \tag{3.4}$$

II) Spin-dependent Gauss-type Afnan–Tang $S_3$ potential [18].

III) Spin-dependent Yukawa-type Malfliet–Tjon I/III potential [19].

$$V_{\text{triplet}}(r)=7.39\frac{\exp[-3.11r]}{r}-3.22\frac{\exp[-1.55r]}{r}\left[fm^{-1}\right],$$
$$V_{\text{singlet}}(r)=7.39\frac{\exp[-3.11r]}{r}-2.64\frac{\exp[-1.55r]}{r}\left[fm^{-1}\right], \tag{3.6}$$

The potential strengths are in MeV for Volkov and Afnan–Tang and dimensionless for Malfliet–Tjon I/III. The range parameters, namely exchanged pion masses are in $fm^{-2}$ for Volkov and Afnan–Tang and $fm^{-1}$ for Malfliet–Tjon I/III potentials. Now, a number of the coupled equations in partial-wave representation are discussed. Previously, in the partial-wave representation of the 3N bound system equations, the number of partial-wave channels had to be high, in order to achieve reasonably well enough converged energy eigenvalues, namely $N_u = 34$, $N_v =$ 20 and $N_\gamma = 5, 18, 26, 34, 42$ ($N_u$ and $N_v$ are the numbers of first and second Jacobi momenta, and $N_\gamma$ is the number of partial-wave channels) [20]. For a realistic calculation in the 3N problem, at least $N_\gamma = 34$ channels are required and thus the dimension of the kernel leads to $N = N_u \times N_v \times N_\gamma \sim 24,000$. In contrast to the 3N system, the number of channels for the 5N and 4N bound state in the case of $\alpha$-states, $N_{S/T} = N_{\gamma_u} + N_{\gamma_v}$ is in principle unlimited where $N_{\gamma u}$ and $N_{\gamma v}$ are the numbers of $\gamma_u$ and $\gamma_v$ quantum number combinations respectively, even if the nucleon–nucleon interaction is assumed to act only in a certain 2N states. We display adequate examples for those maximum $N_{\gamma_u}$ and $N_{\gamma_v}$ values in Table 2. In Table 1, the number of spin-isospin states are presented for both kinds of the Jacobi coordinates, $\gamma_u$ and $\gamma_v$, as well as the number of coupled Yakubovsky equations in our partial-wave representation. The angular momentum quantum numbers do not appear explicitly in our formalism, since the basis is restricted to $L = 1$. Therefore a number of coupled equations which are fixed according to the spin-isospin states are strongly reduced.

## 5. Results

### 5.1. The four- and five-nucleon binding energies

In this section in order to investigate the effective interaction between $\alpha$-particle and an attractive neutron we have presented the numerical results for binding energies of the 5N in the case of $\alpha - n$ structure, and compare with the 4N binding energies as $\alpha$-particle, because the binding energy differences between specific 4N and 5N structures in such a model ($\alpha$-state) refers to the value of effective interaction between $\alpha$-particle and attractive neutron. The results for 4N and 5N bound-state are reported in the below tables, respectively. For spin-independent Volkov potential our numerical results for 4N and 5N binding energies yield the values −30.39 and −44.02 MeV, which as shown in Table 3 are also good agreement with those obtained from other achievements.

For spin-dependent Afnan–Tang potential our numerical results for 4N and 5N binding energies yield the values −28.78 and −44.50 MeV, which as shown in Table 4 are also good compatibility with those achieved from other calculations, namely SVM [7].

Correspondingly, our numerical results for Malfliet–Tjon I/III with the value −30.31 MeV for 4N binding energy, is in good agreement with IDEA [26], EIHH [27] and HH [23].

Table 4
The four- and five-nucleon binding energies for spin-dependent Afnan–Tang potential. The angular momentum quantum numbers are labeled as $(LS)J^\pi$.

| Method | $E_{4N}$ (MeV)$(0\,0)0^+$ | $E_{5N}$(MeV) $\left(1\tfrac{1}{2}\right)\tfrac{3}{2}^-$ |
|---|---|---|
| F-Y [14] | −28.80 | |
| VAR [25] | −25.654 | |
| SVM [7] | −30.37 | −44.27 |
| This work | −28.78 | −44.50 |

Comparisons of our numerical results for binding energies with spin-independent and also spin-dependent nucleon–nucleon type potentials are in reasonable agreement with the obtained of other methods in the 5N Yakubovsky calculations, and the some obtained binding energy differences between 4N ($\alpha$) and 5N ($\alpha - n$) systems, demonstrates that the effective $\alpha N$ interaction attracts to about 13 MeV even for realistic nucleon–nucleon interactions. It is worth to mention that all the scattering state calculations of effective $\alpha N$ interactions with standard methods, confirms the findings of the authors.

### 5.2. Test of the accuracy of the calculations

In order to investigate the accuracy of our numerical solution for the Yakubovsky integral equations (2.11) and (2.12), and also the calculations of the total wave function (1.10), we have calculated the expectation value of the 5N Hamiltonian $\langle H \rangle$ and compared with the calculated binding energy. The explicit form of the expression $\langle H\rangle$ which should be obtained from expectation values of the free Hamiltonian as well as potential are given as

$$\langle H\rangle\equiv\langle\Psi_{\alpha-n}|H|\Psi_{\alpha-n}\rangle=\langle\Psi_{\alpha-n}|H_0|\Psi_{\alpha-n}\rangle+\langle\Psi_{\alpha-n}\left|V_{ij}\right|\Psi_{\alpha-n}\rangle=\left(60\langle\psi^{1234}_{12;123}|H_0|\Psi_{\alpha-n}\rangle+30\langle\psi^{1234}_{12;12+34}|H_0|\Psi_{\alpha-n}\rangle\right)+10\langle\Psi_{\alpha-n}\left|V_{ij}\right|\Psi_{\alpha-n}\rangle, \quad (3.7)$$

the potential term of the expectation value is calculated with Malfliet−Tjon I/III interactions. In Table 6 the obtained eigenvalue results are represented for binding energy $E=-43.90$ MeV for different mesh points. As demonstrated in Table 6, the calculations of the eigenvalue $\eta$ converge to the value one for the number of mesh points Jacobi momenta, angles and spin variables as $N^u_{\rm Jac}=N^v_{\rm Jac}=16$, $N_{\rm Shp}=N_{\rm Azi}=12$ and $N_{S/T}=8$. The comparison between the expectation value of the 5N Hamiltonian $\langle H\rangle$ and the eigenvalue energy $E^{\alpha-n}_{5N}$ shows that our both results are fair in agreement. However, the improved agreement could be achieved if we considered a larger number of mesh points in our calculations.

Table 5
The four- and five-nucleon binding energies for spin-dependent Malfliet–Tjon I/III potential. The angular momentum quantum numbers are labeled as $(LS)J^\pi$.

| Method | $E_{4N}$ (MeV)$(00)0^+$ | $E_{5N}$ (MeV) $\left(1\tfrac{1}{2}\right)\tfrac{3}{2}^-$ |
|---|---|---|
| IDEA [26] | −30.20 | |
| EIHH [27] | −30.71 | |
| HH [23] | −30.33 | |
| This work | −30.31 | −43.90 |

Table 6
Convergence of the eigenvalue $\eta$ of Yakubovky kernel and the expectation value $\langle H \rangle$, with respect to the adequate number of mesh points in Jacobi momenta, angles and spin effects.

| $N^u_{\mathrm{Jac}}$ | $N^v_{\mathrm{Jac}}$ | $N_{\mathrm{Shp}} = N_{\mathrm{Azi}}$ | $N_{S/T}$ | $\eta$ | $\langle H \rangle$ | $E^{\alpha-n}_{5N}$ |
|---|---|---|---|---|---|---|
| 10 | 10 | 12 | 8 | 0.989 | −46.2 | −43.83 |
| 12 | 12 | 12 | 8 | 0.992 | −46.6 | −43.86 |
| 14 | 12 | 12 | 8 | 0.996 | −45.8 | −43.88 |
| 16 | 14 | 12 | 8 | 0.999 | −45.3 | −43.90 |
| 16 | 16 | 12 | 8 | 1.000 | −45.2 | −43.90 |

## 6. Concluding remarks

In this project in order to study the alpha-nucleon as an effective two-body interaction, we have solved the coupled Yakubovsky equations for the 5N and 4N bound state in the case of effective $\alpha$-state structure, like $^5$He and $^4$He nuclei, with respect to the regarded spin and isospin degrees of freedom, which is implemented in the basis of Jacobi momentum representation. First, we formulated the coupled Yakubovsky equations for the 5N in the picture of alpha-neutron structure as the anti-symmetrized function of Jacobi momenta, specially the magnitudes of the momenta and the angles between them. We expect that the coupled integral equations for a bound state can be handled in a partial-wave representation and a numerically reliable standard iteration method. In the calculations, we applied the spin-dependent potential models, *i.e.* Afnan−Tang S3 and Malfliet−Tjon I/III, along with spin-independent potentials. These potentials provide reasonable results for binding energies in comparison with the other results that have been achieved in the previous calculations. In this calculation, as it can be seen, all the obtained results for the 4N binding energies are in excellent compatibility with the other results. Also, the obtained results for 5N system are in fair compatibility with the obtained results from other methods. The calculated binding energies for 4N and 5N bound systems, by three-type potentials with different behaviors, are represented in Tables 3−5. The above-mentioned simplifications naturally make our numerical calculations with deviation, in comparison with the experimental data. However, improved obtained results for the 5N and 4N system in comparison with the spin-independent calculations, with respect to the regarded applying spin-dependent nucleon−nucleon potential models, show that the spin/isospin effects in the calculations play a nontrivial role in approaching the experimental data. In addition, the some obtained binding energy differences between 4N and 5N systems in such structures suggest that the effective interaction of $\alpha N$ occurs to about 13 MeV and is attractive. It is mentioned that all the scattering state calculations of effective $\alpha N$ interactions with standard methods, confirms the findings of the authors.

In order to test the accuracy of the eigenvalue results, the stability of our algorithm and partial-wave representation of Yakubovsky components have been achieved with the calculation eigenvalue of Yakubovsky kernel, where different number of mesh points for Jacobi momenta and angle variables have been used. We have also calculated the expectation value of the 5N Hamiltonian operator. This test of calculation has been done with Malfliet−Tjon I/III potentials and we have achieved a good compatibility between the obtained eigenvalue energy and the expectation value of the Hamiltonian operator.

## Appendix.

*Explicit evaluation of the coupled equations*

To evaluate the coupled equations, Eqs. (2.11) and (2.12), we need to evaluate the below matrix elements in partial-wave analysis

$$\langle u;\gamma_u\left|G_0\mathcal{T}^{123}\right|u';\gamma_{u'}\rangle \quad (A.1)$$

$$\langle u';\gamma_{u'}|P_{34}P_{45}|u'';\gamma_{u''}\rangle \quad (A.2)$$

$$\langle u';\gamma_{u'}|P_{34}|u'';\gamma_{u''}\rangle \quad (A.3)$$

$$\langle u';\gamma_{u'}|v';\gamma_{v'}\rangle \quad (A.4)$$

$$\langle v;\gamma_{v}|G_0\mathscr{T}^{12+34}|v';\gamma_{v'}\rangle \quad (A.5)$$

$$\langle v';\gamma_{v'}|u';\gamma_{u'}\rangle \quad (A.6)$$

$$\langle v';\gamma_{v'}|P_{45}|u';\gamma_{u'}\rangle \quad (A.7)$$

$$\langle v';\gamma_{v'}|P_{34}|u';\gamma_{u'}\rangle \quad (A.8)$$

$$\langle v';\gamma_{v'}|P_{34}P_{45}|u';\gamma_{u'}\rangle \quad (A.9)$$

To evaluate the first term, Eq. (A.1), we need to solve the first sub cluster Faddeev-like equation to obtain $\mathscr{T}^{123}$ by using Pade' approximation [14], as follows

$$G_0\mathscr{T}^{123} = G_0t_{12}P + G_0t_{12}PG_0t_{12}P + G_0t_{12}PG_0t_{12}PG_0t_{12}P + \cdots. \quad (A.10)$$

The basis states that used in Eqs. (2.11) and (2.12) are restricted in low-lying states, namely without angular momentum quantum numbers, though, we have considered the angular grid point in Section 4 as selecting the suitable coordinate system. Therefore, for such a basis states the C.G. coefficients are generally defined in a standard PW analysis

$$\langle\gamma_v|\gamma_u\rangle = G_{uv} = \delta_{T\gamma_v T\gamma_u}\delta_{M_{T_v}M_{T_u}}(-1)^{s_{12}+I_3+1}\sqrt{\widehat{j_2}\widehat{s_{23}}\widehat{I_3}\widehat{s_{34}}\widehat{s_5}\widehat{s_{45}}}\begin{Bmatrix} s_{12} & \frac{1}{2} & I_3 \\ \frac{1}{2} & S & s_{23}\end{Bmatrix}\begin{Bmatrix} I_3 & \frac{1}{2} & S \\ \frac{1}{2} & \frac{1}{2} & s_{45} \\ I_4 & s_{45} & J\end{Bmatrix} \times(-1)^{t_{12}+t_4+1}\sqrt{\widehat{t_2}\widehat{t_{23}}\widehat{t_4}\widehat{t_{45}}}\begin{Bmatrix} t_{12} & \frac{1}{2} & t_3 \\ \frac{1}{2} & t_4 & t_{23}\end{Bmatrix}\begin{Bmatrix} t_3 & \frac{1}{2} & t_4 \\ \frac{1}{2} & \frac{1}{2} & t_{45} \\ t_4 & t_{45} & T\end{Bmatrix}, \quad (A.11)$$

hence, the $G_{uv}$ in each bellow term is the corresponding geometrical coefficient. To evaluate the first term of Eq. (A.10), we should again insert a completeness relation between the 2N $t$-matrix and permutation operator $P \equiv P_{12}P_{23} + P_{13}P_{23}$ [14] as

$$\langle u;\gamma_u|G_0\mathscr{T}^{123}|u';\gamma_{u'}\rangle = G_0\int u''^2 Du''\langle u;\gamma_u|t_{12}|u'';\gamma_{u''}\rangle \times\langle u'';\gamma_{u''}|P|u';\gamma_{u'}\rangle, \quad (A.12)$$

where

$$\langle u;\gamma_u|t_{12}|u'';\gamma_{u''}\rangle = \langle\gamma_u|\gamma_{u''}\rangle\langle u_1|t_{12}|u_1''\rangle\langle u_2|u_2''\rangle\langle u_3|u_3''\rangle \times\langle u_4|u_4''\rangle, \quad (A.13)$$

and

$$\langle u'';\gamma_{u''}|P|u';\gamma_{u'}\rangle = \langle\gamma_u|\gamma_{u''}\rangle\langle u'|P_{12}P_{23}|u''\rangle + \langle\gamma_u|\gamma_{u''}\rangle \times\langle u'|P_{13}P_{23}|u''\rangle \quad (A.14)$$

and

$$\langle u;\gamma_u|t_{12}|u'';\gamma_{u''}\rangle = \langle u_1|t_{(\varepsilon)}|u_1''\rangle G_{uu''}\frac{\delta(u_2-u_2'')}{(u_2'')^2}\frac{\delta(u_3-u_3'')}{(u_3'')^2}\frac{\delta(u_4-u_4'')}{(u_4'')^2};\ \ \varepsilon = E-\frac{3}{4}\frac{u_2^2}{m}-\frac{2}{3}\frac{u_3^2}{m}-\frac{5}{8}\frac{u_4^2}{m}, \tag{A.15}$$

here $\varepsilon$ is the energy of 2N subsystem in $u$-set (see Eq. (2.5)), and

$$\langle u;\gamma_u|P|u'';\gamma_{u''}\rangle = \frac{\delta(u_3''-u_3')}{(u_3')^2}\frac{\delta(u_4''-u_4')}{(u_4')^2}G_{u''u'}\int_{-1}^{1} du_{2''2'}\frac{\delta[u_1'-|-\frac{1}{2}\mathbf{u}_2''-\mathbf{u}_2'|]}{[u_1']^2}\frac{\delta[u_1''-|\frac{1}{2}\mathbf{u}_2'+\mathbf{u}_2''|]}{[u_1'']^2}. \tag{A.16}$$

To evaluate the term of Eq. (A.2), there is a relation between different basis states in sub-clusters $(123+4+5;12)$ and $(125+3+4;12)$,

$$\langle u';\gamma_{u'}|P_{34}P_{45}|u'';\gamma_{u''}\rangle = \frac{1}{2^3}\frac{\delta(u_1'-u_1'')}{(u_1'')^2}G_{u'u''}\int_{-1}^{1} du_{23}\int_{-1}^{1} du_{24}\int_{-1}^{1} du_{34}$$
$$\times\frac{\delta[u_2'-|\frac{1}{3}\mathbf{u}_2''+\frac{2}{9}\mathbf{u}_3''+\frac{5}{12}\mathbf{u}_4''|]}{(u_2'')^2}\frac{\delta[u_3''-|\mathbf{u}_2''+\frac{1}{12}\mathbf{u}_3''-\frac{5}{16}\mathbf{u}_4''|]}{(u_3'')^2}\frac{\delta[u_4''-|\mathbf{u}_3''-\frac{1}{5}\mathbf{u}_4''|]}{(u_4'')^2}, \tag{A.17}$$

To evaluate the term of Eq. (A.3), there is a relation between different basis states in sub-clusters $(123+4+5;12)$ and $(124+3+5;12)$,

$$\langle u';\gamma_{u'}|P_{34}|u'';\gamma_{u''}\rangle = \frac{1}{2}\frac{\delta(u_1'-u_1'')}{(u_1'')^2}\frac{\delta(u_4'-u_4'')}{(u_4'')^2}G_{u'u''}\int_{-1}^{1} du_{23}\frac{\delta[u_2''-|\frac{1}{3}\mathbf{u}_2''+\frac{8}{9}\mathbf{u}_3''|]}{(u_2'')^2}\frac{\delta[u_3''-|\mathbf{u}_2''-\frac{1}{3}\mathbf{u}_3''|]}{(u_3'')^2}. \tag{A.18}$$

To evaluate the term of Eq. (A.4), there is a relation between different basis states in sub-clusters $(123+4+5;12)$ and $(12+34+5;12)$,

$$\langle u';\gamma_{u'}|v';\gamma_{v'}\rangle=\frac{1}{2}\frac{\delta(u'_1-v'_1)}{(v'_1)^2}\frac{\delta(u'_4-v'_4)}{(v'_4)^2}G_{u'v'}\int_{-1}^{1}\mathrm{d}u_{23}\frac{\delta[v'_2-|\frac{1}{2}\boldsymbol{u}''_2-\frac{2}{3}\boldsymbol{u}''_3|]}{(v'_2)^2}\frac{\delta[v'_3-|\boldsymbol{u}''_2-\frac{2}{3}\boldsymbol{u}''_3|]}{(v'_3)^2}, \tag{A.19}$$

Correspondingly, to evaluate Eq. (A.5) we need to solve the sub-cluster Faddeev-like equation to obtain $\mathscr{T}^{12+34}$ by using Pade' approximation [14], as follows

$$G_0\mathscr{T}^{12+34}=G_0t_{12}\tilde{P}+G_0t_{12}\tilde{P}G_0t_{12}\tilde{P}+G_0t_{12}\tilde{P}G_0t_{12}\tilde{P}G_0t_{12}\tilde{P}+\cdots. \tag{A.20}$$

To evaluate the first term of Eq. (A.20) we should insert again a completeness relation between the 2N $t$-matrix operator and permutation operator $\tilde{P}\equiv P_{13}P_{24}$ [14] as

$$\langle v;\gamma_v|G_0\mathscr{T}^{12+34}|v';\gamma_{v'}\rangle=G_0\int v''^2Dv''\langle v;\gamma_v|t_{12}|v'';\gamma_{v''}\rangle\times\langle v'';\gamma_{v''}|\tilde{P}|v';\gamma_{v'}\rangle, \tag{A.21}$$

where

$$\langle v;\gamma_v|t_{12}|v'';\gamma_{v''}\rangle=\langle\gamma_v|\gamma_{v''}\rangle\langle v_1|t_{12}|v''_1\rangle\langle v_2|v''_2\rangle\langle v_3|v''_3\rangle\times\langle v_4|v''_4\rangle, \tag{A.22}$$

and to evaluate the matrix elements of permutation operator $\tilde{P}=P_{13}P_{24}$ [13] we have used the relation between different basis states in $(34+12+5;12)$ and $(12+34+5;12)$

$$\langle v'';\gamma_{v''}|\tilde{P}|v';\gamma_{v'}\rangle=\langle v'';\gamma_{v''}|P_{13}P_{24}|v';\gamma_{v'}\rangle=G_{v''v'}\frac{\delta(v''_1-v'_2)}{(v'_2)^2}\frac{\delta(v''_2-v'_1)}{(v'_1)^2}\frac{\delta(v''_3-v'_3)}{(v'_3)^2}\frac{\delta(v''_4-v'_4)}{(v'_4)^2}, \tag{A.23}$$

the matrix elements of 2N $t$-matrix and permutation operator $\tilde{P}$ are evaluated

$$\langle v;\gamma_v|t_{12}|v'';\gamma_{v''}\rangle=\langle v_1|t_{(\varepsilon^*)}|v''_1\rangle G_{v'v''}\frac{\delta(v''_2-v_2)}{(v''_2)^2}\frac{\delta(v''_3-v_3)}{(v''_3)^2}\frac{\delta(v''_4-v_4)}{(v''_4)^2};\ \varepsilon^*=E-\frac{v_2^2}{m}-\frac{1}{2}\frac{v_3^2}{m}-\frac{5}{8}\frac{v_4^2}{m}, \tag{A.24}$$

where $\varepsilon^*$ is the energy of 2N subsystem in $v$-set (see Eq. (2.7)). To evaluate Eq. (A.6) we should use the relation between different basis states in $(12+34+5;12)$ and $(123+4+5;12)$,

$$\langle v';\gamma_{v'}|u';\gamma_{u'}\rangle=\frac{1}{2}\frac{\delta(v'_1-u'_1)}{(u'_1)^2}\frac{\delta(v'_4-u'_4)}{(u'_4)^2}G_{v'u'}\times\int_{-1}^{1}\mathrm{d}v_{32'}\frac{\delta[v'_2-|\frac{2}{3}\boldsymbol{v}'_2-\frac{2}{3}\boldsymbol{v}'_3|]}{(v'_2)^2}\frac{\delta[v'_3-|-\boldsymbol{v}'_2-\frac{1}{2}\boldsymbol{v}'_3|]}{(v'_3)^2}. \tag{A.25}$$

To evaluate the term of Eq. (A.7) we should use the relation between different basis states in $(12+34+5;12)$ and $(123+5+4;12)$,

$$\langle v';\gamma_{v'}|P_{45}|u';\gamma_{u'}\rangle=\frac{1}{2^3}\frac{\delta(u'_1-v'_1)}{(u'_1)^2}\frac{\delta[u'_2-v'_2]}{(u'_2)^2}G_{v'u'}\times\int_{-1}^{1}\mathrm{d}v_{32'}\int_{-1}^{1}\mathrm{d}v_{42'}\int_{-1}^{1}\mathrm{d}v_{34}\times\frac{\delta[u'_3-|-\frac{1}{4}\boldsymbol{v}'_2-\frac{1}{8}\boldsymbol{v}'_3+\frac{3}{4}\boldsymbol{v}'_4|]}{(u'_3)^2}\times\frac{\delta[u'_4-|-\frac{1}{2}\boldsymbol{v}'_2-\frac{1}{4}\boldsymbol{v}'_3+\frac{1}{2}\boldsymbol{v}'_4|]}{(u'_4)^2}. \tag{A.26}$$

To evaluate the term of Eq. (A.8), we should use the relation between different basis states in $(12+34+5;12)$ and $(123+5+4;12)$,

$$\langle v';\gamma_{v'}|P_{34}|u';\gamma_{u'}\rangle = \frac{1}{2}\frac{\delta(u'_1-v'_1)}{(u'_1)^2}\frac{\delta(u'_4-v'_4)}{(u'_4)^2}G_{v'u'}\int_{-1}^{1}\mathrm{d}v_{32'}\frac{\delta\left[u'_2-\left|-\frac{2}{3}\boldsymbol{v}'_2-\frac{2}{3}\boldsymbol{v}'_2\right|\right]}{(u'_2)^2}\frac{\delta\left[u'_3-\left|\boldsymbol{v}'_2-\frac{1}{2}\boldsymbol{v}'_3\right|\right]}{(u'_3)^2}. \tag{A.27}$$

To evaluate the fourth term of Eq. (A.9) we should use the relation between different basis states in $(12+34+5;12)$ and $(124+5+3;12)$,

$$\langle v';\gamma_{v'}|P_{34}P_{45}|u';\gamma_{u'}\rangle = \frac{1}{2^3}\frac{\delta(u'_1-v'_1)}{(u'_1)^2}G_{v'u'}\int_{-1}^{1}\mathrm{d}v_{32'}\int_{-1}^{1}\mathrm{d}v_{42'}\int_{-1}^{1}\mathrm{d}v_{34} \times\frac{\delta\left[u'_2-\left|-\frac{2}{3}\boldsymbol{v}'_2-\frac{2}{3}\boldsymbol{v}'_2\right|\right]}{(u'_2)^2}\frac{\delta\left[u'_3-\left|\frac{1}{4}\boldsymbol{v}'_2-\frac{1}{8}\boldsymbol{v}'_3+\frac{3}{4}\boldsymbol{v}'_4\right|\right]}{(u'_3)^2}\frac{\delta\left[u'_4-\left|\frac{1}{2}\boldsymbol{v}'_2-\frac{1}{4}\boldsymbol{v}'_3+\frac{1}{2}\boldsymbol{v}'_4\right|\right]}{(u'_4)^2}. \tag{A.28}$$

In the above-mentioned integral factors the quantity $u_{ij}$ $(v_{ij})$ is related to the angle variable between $\boldsymbol{u}_i$ and $\boldsymbol{u}_j$ ($\boldsymbol{v}_i$ and $\boldsymbol{v}_j$), *i.e.* $u_{ij}\equiv\cos(\boldsymbol{u}_i\boldsymbol{u}_j)$ and $v_{ij}\equiv\cos(\boldsymbol{v}_i,\boldsymbol{v}_j)$, respectively (More discussions in Section 4).